\documentclass[twocolumn,aps,pre,footinbib,floatfix,preprintnumbers,amsmath,amssymb,superscriptaddress,10pt,colorlinks]{revtex4-1}

\usepackage{ulem}
\renewcommand{\emph}{\textit}

\usepackage{selinput}
\usepackage{amsmath}
\usepackage{amssymb}
\usepackage{graphicx}
\usepackage{bm} 
\usepackage[usenames,dvipsnames]{color}
\usepackage{natbib}
\usepackage{dsfont}
\usepackage{xr}
\usepackage{mathtools}
\usepackage{here}
\usepackage[urlcolor=RoyalBlue,citecolor=RoyalBlue,linkcolor=BrickRed]{hyperref}
\usepackage{tensor}
\usepackage{textcomp}

\makeatletter
\newcommand{\annotate}[2][]{%
\pdfstringdef\x@title{#1}%
\edef\r{\string\r}%
\pdfstringdef\x@contents{#2}%
\pdfannot
width 2\baselineskip
height 2\baselineskip
depth 0pt
{
/Subtype /Text
/T (\x@title)
/Contents (\x@contents)
}%
}
\makeatother

\newcommand{\mean}[1]{\left < #1 \right >}
\newcommand{\abs}[1]{\left | #1 \right |}

\renewcommand{\vec}[1]{\mathbf{ #1 }}

\newcommand{\red}[1]{\textcolor{BrickRed}{#1}}

\newcommand{\para}{{\mkern3mu\vphantom{\perp}\vrule depth 0pt\mkern2mu\vrule depth 0pt\mkern3mu}}

\makeatletter
\newcommand*{\balancecolsandclearpage}{%
  \close@column@grid
  \cleardoublepage
  \twocolumngrid
}
\makeatother

\begin{document}

\title{A particle-field representation unifies paradigms in active matter}

\author{Robert Gro{\ss}mann}
\email{grossmann@physik.hu-berlin.de}
\affiliation{Laboratoire J.A.~Dieudonn\'{e}, Universit\'{e} C\^{o}te d'Azur, UMR 7351 CNRS, 06108 Nice, France}
\affiliation{Institute of Physics and Astronomy, University of Potsdam, D-14476 Potsdam, Germany}

\author{Igor S. Aranson}
\email{isa12@psu.edu}
\affiliation{Department of Biomedical Engineering, Pennsylvania State University, University Park, Pennsylvania, 16802, USA}

\author{Fernando Peruani}
\email{peruani@unice.fr}
\affiliation{Laboratoire J.A.~Dieudonn\'{e}, Universit\'{e} C\^{o}te d'Azur, UMR 7351 CNRS, 06108 Nice, France}

\begin{abstract}
Active matter research focuses on the emergent behavior among interacting self-propelled particles. 
Unification of seemingly disconnected paradigms~--~active phase-separation of repulsive discs and collective motion of self-propelled rods~--~is a major challenge in contemporary active matter. 
Inspired by the quanto-mechanical wave-particle duality, we develop an  approach based on the representation of active particles by smoothed continuum fields. 
On the basis of the collision kinetics, we demonstrate analytically and numerically how nonequilibrium stresses acting among self-driven, anisotropic objects hinder the formation of phase-separated states as observed for self-propelled discs and facilitate the emergence of orientational order. 
Besides particle shape, the rigidity of self-propelled objects controlling the symmetry of emergent ordered states is as a crucial parameter:~impenetrable, anisotropic rods are found to form polar, moving clusters, whereas large-scale nematic structures emerge for soft rods, notably separated by a bistable coexistence regime. 
These results indicate that the symmetry of the ordered state is not dictated by the symmetry of the interaction potential but is rather a dynamical, emergent property of active systems. 
This unifying theoretical framework can represent a variety of active systems:~living cell tissues, bacterial colonies, cytoskeletal extracts as well as shaken granular media. 
\end{abstract}

\date{\today}
\maketitle


\section{Introduction}

Interacting self-propelled particles, termed active matter, are the standard model of collective behavior out of thermodynamic equilibrium~\cite{romanczuk_active_2012,marchetti_hydrodynamics_2013,menzel_tuned_2015}. 
Active systems become increasingly popular in different disciplines studying diverse systems from micron-sized synthetic swimmers~\cite{paxton_catalytic_2004,palacci_living_2013,theurkauff_dynamic_2012} or phoretic colloids~\cite{buttinoni_dynamical_2013, golestanian_designing_2007}, collective swimming and self-organization of bacteria~\cite{dombrowski_self-concentration_2004, sokolov_concentration_2007} as well as self-assembly in biomimetic systems~\cite{sanchez_spontaneous_2012} to the behavior of insect swarms~\cite{attanasi_finite_2014}, sheep herds~\cite{ginelli_intermittent_2015}, fish schools~\cite{ward_quorum_2008} and flocks of birds~\cite{ballerini_interaction_2008}.

The current theoretical understanding of active matter is based on two cornerstones. 
One of these paradigms is the emergence of phase-separated states in ensembles of self-propelled discs~\cite{fily_athermal_2012,redner_structure_2013,digregorio_full_2018}, also referred to as active Brownian particles~\footnote{The term \textit{active Brownian particles}, originally introduced in~\cite{schimansky_structure_1995} referring to Brownian particles with the ability to generate a field which influences their motion, is often used synonymously in a more general context with self-propelled motion far from equilibrium~\cite{schweitzer_complex_1998,romanczuk_active_2012}.}, due to the combined effect of self-propulsion and isotropic repulsion~\cite{cates_mips_2015}.
This phenomenon, sharing similarities with granular media agitated by vibration~\cite{aranson_patterns_2006}, was termed \textit{motility-induced phase separation}. 
Its theoretical appeal stems from the potential mapping of the nonequilibrium dynamics at large scales to an effective equilibrium theory for the density field~\cite{solon_pressure2_2015,nardini_entropy_2017,solon_generalized_2018,solon_generalized2_2018,prymidis2016,prymidis2017,bialke_microscopic_2013,haertel_three_2018,haertel_three_2018}. 
%
%
Despite various experimental realizations of self-propelled discs were designed~\cite{buttinoni_dynamical_2013,deseigne_collective_2010}, experimental evidence of this very type of active phase separation is still lacking. 
Recent experiments with active Janus colloids suggest the hypothesis that polar orientational ordering within clusters interrupts motility-induced phase separation~\cite{vanlinden_interrupted_2019}. 
%

Orientational symmetry breaking and the emergence of collective motion due to velocity alignment is another central paradigm in active matter~--~the exploration of phases with orientational order is an integral part of it~\cite{vicsek_novel_1995,toner_long_1995,chate_simple_2006,chate_modeling_2008,ginelli_large_2010,ramaswamy_mechanics_2010}.\footnotetext{Velocity-alignment may be induced further by hydrodynamic interactions~\cite{hoell_particle_2018} or dissipative collisions~\cite{grossman_emergence_2008}. We note that torques are typically assumed to be absent for isotropic self-propelled discs. However, torques do only vanish for isotropic, self-propelled particles only if tangential friction among particles and hydrodynamic flows are negligible~\cite{hoell_particle_2018}. }
The most relevant source of alignment is anisotropic repulsion of spatially extended, self-propelled objects~\cite{peruani_nonequilibrium_2006,Note1}.
%
%
Recent numerical studies unveiled a large variety of collective phenomena among self-propelled, anisotropic objects including mesoscale-turbulence~\cite{wensink_emergent_2012}, formation of bands and aggregates~\cite{weitz_selfpropelled_2015}, accumulation at confining walls~\cite{wensink_aggregation_2008} and a complex phase diagram depending on the rigidity~\cite{abkenar_collective_2013,shi_self_2018} and deformability of the particles~\cite{menzel2012soft,ohta2014traveling,ohta2014traveling,tarama_individual_2014,lober_collisions_2015}. 
Beyond the inherent theoretical interest in the physics of self-propelled particles, there exists a large number of real-world applications:~motile bacteria in two-dimensions~\cite{peruani_collective_2012,wensink_meso-scale_2012,zhang_collective_2010,sokolov_concentration_2007}, biomimetic systems such as motility assays~\cite{schaller_polar_2010,sanchez_spontaneous_2012,sumino_large-scale_2012,tanida_gliding_2018,huber_emergence_2018} as well as shaken~\cite{narayan_long_2007,aranson_swirling_2007} and self-propelled, granular rods~\cite{kudrolli_swarming_2008}.

In a nutshell:~particle shape controls the physics of active systems~--~whereas active phase separation is expected for self-propelled discs, anisotropic volume exclusion leads to collectively moving clusters~\cite{peruani_nonequilibrium_2006}.
Based on that insight, a unified framework encompassing central phenomena observed in active systems can be established as the basis for a theory of active matter. 
In this context, novel theoretical concepts are called for as the application of concepts from equilibrium statistical mechanics to active matter is limited to a few special cases~\cite{solon_pressure_2015,fodor_how_2016}.

The complexity of models for spatially extended, anisotropic objects has hindered analytical studies and systematic coarse-graining addressing their collective properties~--~the characterization has been mainly carried out mainly by numerical simulations~\footnote{Clustering properties were studied in terms of phenomenological coagulation-fragmentation equations~\cite{peruani_nonequilibrium_2006,peruani_kinetic_2013}}. 
The derivation of hydrodynamic equations from microscopic models has only been possible for heuristic point particles with a prescribed velocity-alignment rule based on symmetry considerations and binary collisions~\cite{aranson_pattern_2005,aranson_patterns_2006, bertin_boltzmann_2006,ihle_kinetic_2011,peshkov_nonlinear_2012,peshkov_boltzmann_2014,grossmann_pattern_2015,heidenreich_hydrodynamic_2016}.

Inspired by the quanto-mechanical wave-particle duality, we develop in this work a unifying modeling approach for soft, self-driven agents: each individual entity is represented by an anisotropic, smoothed field whose mutual interactions are derived from the minimization of an overlap energy.
Thus, the interaction force and the torque result from a single interaction principle governed by the anisotropy of particles. 
This approach yields an universal, simple and descriptive model that can be analytically coarse-grained, unifying different paradigms of active matter:~motility-induced phase separation and the emergence of orientational order. 
Combining numerical simulations and analytically derived coarse-grained order parameter equations, we show that motility-induced phase separation cannot emerge for sufficiently anisotropic objects due to the combined action of self-propulsion and anisotropic repulsion. 
The resulting nonequilibrium stresses acting on the microscale  induce orientational alignment of different symmetries locally. 
In this system, the rigidity of particles determines the symmetry of ordered states: long-lived, giant moving clusters are observed if particles strongly repel each other to prevent particle crossing, whereas large-scale nematic order emerges for particles which can slide over each over.
We further indicate that those regimes are separated by a bistable coexistence region, similar to the recently reported ones in motility assays~\cite{huber_emergence_2018}. 
In short, this study sheds light on the importance of anisotropic repulsion as a source of orientational alignment, particularly on how the interrelation of particle shape, rigidity and self-propulsion determines emergent collective behavior~--~key elements to be considered in the design of biomimetic materials.
Unifying different paradigms of active matter within one theoretical framework is expected to pave the way towards a systematic understanding of soft and deformable active matter such as living cell tissues~\cite{lober_collisions_2015}, bacterial colonies~\cite{beer_statistical_2019} and driven filaments~\cite{huber_emergence_2018,kumar_tunable_2018}.

\section{Particle-field representation}
\label{sec:model}

 \begin{figure*}[t]
    \begin{minipage}{0.49\textwidth}
	        \vspace*{0.1cm}
 		\includegraphics[width=0.98\columnwidth]{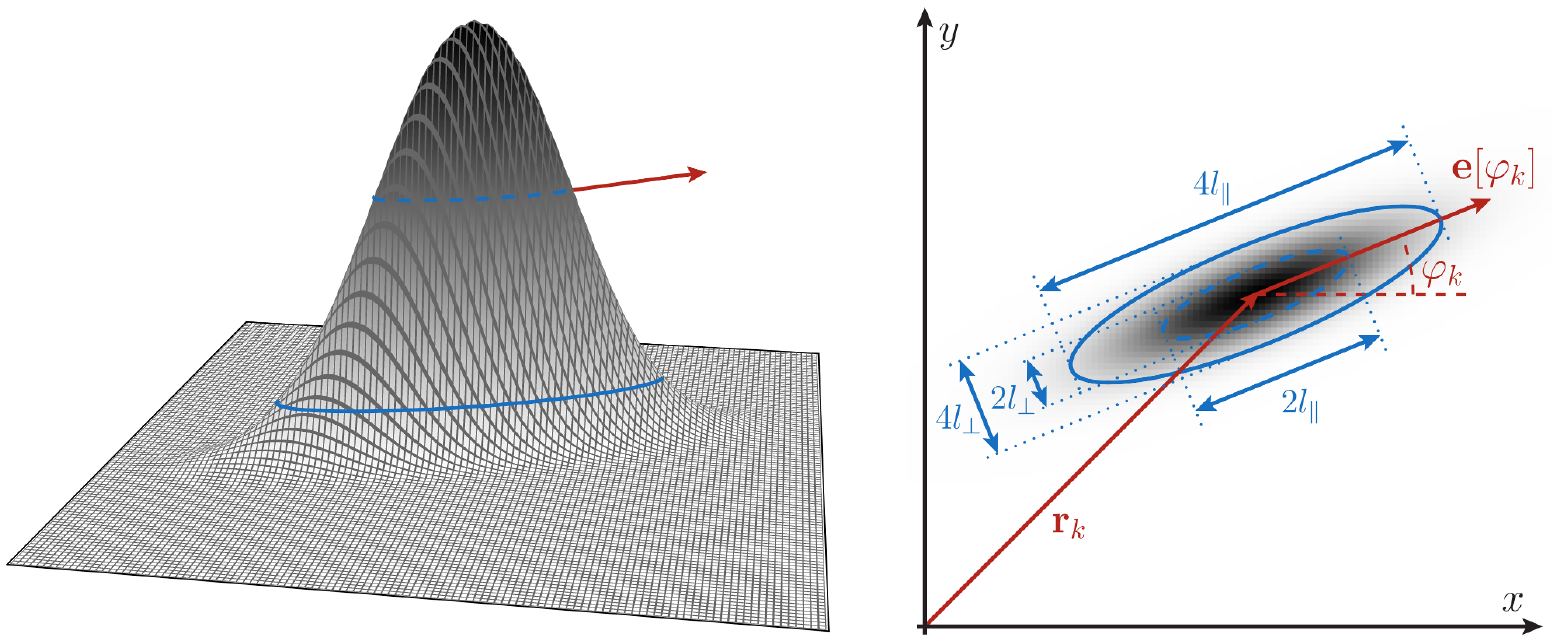}
    \end{minipage}
    \begin{minipage}{0.49\textwidth}
		\includegraphics[width=0.49\columnwidth]{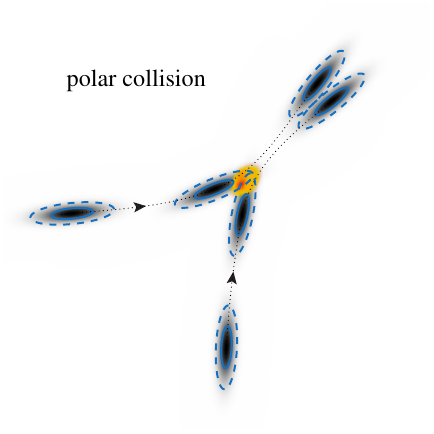}
		\hspace*{-0.7cm}
		\includegraphics[width=0.49\columnwidth]{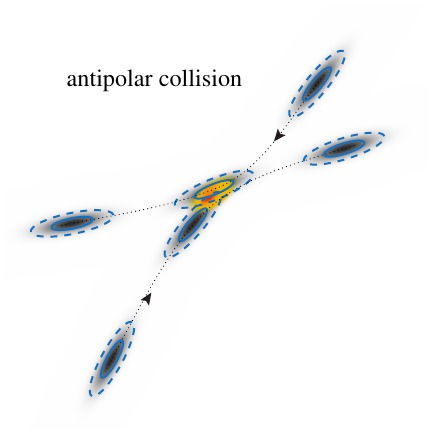}
    \end{minipage}
	\vspace*{-0.3cm}
	\caption{Illustration of a smoothed particle~$\psi \! \left( \vec{r}; \vec{r}_{k},\! \varphi_{k} \right)$, cf.~Eq.~\eqref{eqn:GaussianDensity2}, highlighting the geometric quantities~$l_{\parallel}$, $l_{\perp}$, the orientations~$\vec{e}_{\parallel,\perp} \! \left[ \varphi_{k} \right]$ and the center of mass position~$\vec{r}_{k}$.  Here $l_{\parallel}/l_{\perp} = 4$, $\varepsilon = 15/17$. On the right, a polar and an antipolar collision of rigid rods is illustrated. The overlap upon collision is highlighted in color. After a polar collision, rods apparently tend to move in parallel such that their positions and orientations are highly correlated. In contrast, particles are quickly separated far from another after an antipolar collision. Therefore, the probability to find two rods moving in parallel is enhanced, and polar collisions are the precursor for clustering. 
	} 
 	\label{fig:RodEx}
 \end{figure*}

Inspired by the wave-particle duality, we represent an active particle at position~$\vec{r}_k$ and orientation~$\varphi_k$ by an anisotropic, Gaussian distribution $\psi_k \! \left( \vec{r} \right)$ -- see Fig.~\ref{fig:RodEx} -- as follows: 
\begin{align}
\label{eqn:GaussianDensity2}
\psi_k \! \left( \vec{r} \right) = \!\: e^{ - \frac{ \left \{ (\vec{r}-\vec{r}_{k}) \cdot \vec{e}_{\!\!\:\para} \! \left [ \varphi_k \right ] \right \}^2 }{2 l_{\! \para}^2} - \frac{\left \{ (\vec{r}-\vec{r}_{k}) \cdot \vec{e}_{\!\!\:\perp} \! \left [ \varphi_k \right ] \right \}^2 }{2 l_{\! \perp}^2} } . 
\end{align}
This type of representation is reminiscent  of the procedure used in smoothed-particle hydrodynamics~\cite{gingold_smoothed_1977,lucy_numerical_1977,monaghan_smoothed_2005} as well as Gaussian model potentials to describe molecular interactions~\cite{berne_gaussian_1972}. 
In Eq.~(\ref{eqn:GaussianDensity2}), the dimensions of the major and minor axis are parametrized by~$l_{\para}$ and~$l_{\perp}$, respectively, which are oriented along~$\vec{e}_{\!\!\:\para} \! \left [ \varphi \right ] = \left( \cos \varphi, \sin \varphi \right)$ and $\vec{e}_{\!\!\:\perp} \! \left [ \varphi \right ] = \left( -\sin \varphi, \cos \varphi \right)$.
Throughout, the anisotropy 
\begin{align}
	\varepsilon = \frac{l_{\para}^2 - l_{\perp}^2}{l_{\para}^2 + l_{\perp}^2} 
\end{align}
or, the aspect ratio~$l_{\para} / l_{\perp}$ is the central control parameter for the particle shape determining  the collective properties as well. Particles are circular for~$\varepsilon \!=\! 0$ ($l_{\para} / l_{\perp} = 1$) and  needle-shaped in the limit~$\varepsilon \! \rightarrow \! 1$ ($l_{\para} / l_{\perp}\! \rightarrow \! \infty$).

\paragraph*{Interaction energy, force and torque}
\label{para;inteng}

The basic idea is that particles repel each other to minimize their mutual overlap upon encounter~\cite{berne_gaussian_1972,peruani_nonequilibrium_2006,lober_collisions_2015}. 
For the Gaussian representation of individual active particles, the overlap~$\mathcal{I}_{kj} \!\!=\!\! \int \! d^2 r \, \psi_k \! \left( \vec{r} \right) \psi_j \! \left( \vec{r} \right)$ can be calculated analytically,
\begin{align}
	\mathcal{I}_{kj} \propto e^{ - \frac{ \left( \vec{r}_k - \vec{r}_j \right) \cdot \left [ \mathds{1} - \frac{\varepsilon}{2} \left( \mathcal{Q}\left [ \varphi_k \right ] + \mathcal{Q}\left [\varphi_j \right ] \right) \right ] \cdot \left( \vec{r}_k - \vec{r}_j \right) }{2 \left [ 1 - \varepsilon^2 \cos^2 \!\!\: \left( \varphi_k - \varphi_j \right) \right ] \! \left ( l_{\para}^2 + l_{\perp}^2 \right )} } ,
\end{align}
where~$\mathcal{Q} = \vec{e}_{\!\!\:\para} \otimes \vec{e}_{\!\!\:\para} - \vec{e}_{\!\!\:\perp} \otimes \vec{e}_{\!\!\:\perp}$~(see Supplemental Material~(SM) for technical details of the calculation~\cite{SI}). 
The interaction energy~$\mathcal{U}$ is defined as the sum of binary contributions:~$\mathcal{U} = \frac{1}{2} \sum_{k,j}^N u_2 \! \left( \vec{r}_{k}-\vec{r}_{j}; \varphi_k,\varphi_j \right)$, where  the binary energy~$u_2$ is  an increasing function of the overlap~$\mathcal{I}_{kj}$, 
\begin{align}
	\label{eqn:epot_u2}
	u_2 \! \left( \Delta \vec{r}; \varphi,\varphi' \right) = \kappa \mathcal{F} \!\left[ e^{ - \frac{ \Delta \vec{r} \cdot \left [ \mathds{1} - \frac{\varepsilon}{2} \left( \mathcal{Q}\left [ \varphi \right ] + \mathcal{Q}\left [\varphi' \right ] \right) \right ] \cdot \Delta \vec{r} }{2 \left [ 1 - \varepsilon^2 \cos^2 \!\!\: \left( \varphi - \varphi' \right) \right ] \!   \left ( l_{\para}^2 + l_{\perp}^2 \right )} } \right] \! , 
\end{align}
where $\Delta \vec{r} = \vec{r}'-\vec{r}$ is the relative position, $\kappa$ is the interaction strength measured in units of energy and $\mathcal{F}[\xi]$ is a monotonically increasing function of the overlap. 
This model fulfills two consistency requirements. First, the energy is minimized if particles are nematically~(uniaxially) aligned and is therefore in line with Onsager's mean-field theory~\cite{onsager1949}. 
Second, the energy increases as particles approach each other, hence inducing a repulsive force. 
In particular, soft and hard objects can be described:~if the energy is finite for~$\Delta \vec{r} \rightarrow 0$, particles are soft whereas these objects can be considered hard if the energy diverges in this limit.

The binary force~$\vec{f}_{2}\! \left( \vec{r},\varphi,\varphi' \right) \!=\! - \nabla u_2\! \left( \vec{r},\varphi,\varphi' \right)$ and torque~$m_{2}\! \left( \vec{r},\varphi,\varphi' \right) \!=\! -\partial_{\varphi} u_2\! \left( \vec{r},\varphi,\varphi' \right)$ which a particle at the origin with orientation~$\varphi'$ exerts on a particle at position~$\vec{r}$ and orientation~$\varphi$ are deduced from the potential energy by differentiation with respect to its position and orientation, respectively~(cf.~SM~\cite{SI}). 
The calculation  yields an anisotropic repulsive force between the centers of mass of two particles. 
Newton's third law of \textit{action-reaction} holds for this model. Further, the torque consists of two contributions: 
\begin{align}
	\label{eqn:bin_torques}
	m_2\!  \left( \vec{r},\varphi,\varphi' \right) = \,  & A\!  \left( \vec{r},\varphi,\varphi' \right) \sin \! \big [ 2 \! \left( \varphi - \arg(\vec{r}) \right) \!\!\:\big ] \\
	+ &  B\!  \left( \vec{r},\varphi,\varphi' \right) \sin \! \big [ 2 \! \left( \varphi' - \varphi \right) \! \big ].  \nonumber 
\end{align}
The first one couples the orientation of a rod to the relative position of another rod:~if a rod approaches the center of mass of its interaction partner, it will turn away, thereby minimizing the relative overlap. 
We refer to this term as \textit{collision avoidance}. 
The second contribution is \textit{nematic alignment} of the body axes~\cite{peruani_mean_2008,ginelli_large_2010}. 
The prefactors~$A$ and $B$ are given in the appendix~\cite{SI}.

We note that forces and torques scale differently in terms of the anisotropy:~the force is a zeroth order effect, collision avoidance is a first order effect in the anisotropy~$\varepsilon$, whereas nematic alignment scales with~$\varepsilon^2$ consistently. 
Accordingly, the torque vanishes identically and the force reduces to an isotropic central body force as expected for spherical particles~($\varepsilon = 0$).

Writing force and torque in an analytically closed form~(cf.~\cite{SI}) in contrast to rule-based algorithms is advantageous in several regards. 
The shape of particles is implicit in the force and torque in this approach. 
It is thus not necessary to recalculate overlaps or integrate over the cell area  in every time step, thereby allowing for a fast numerical implementation. Furthermore, smoothed representations of particles are much easier to handle than hard rods with volume exclusion, reflected by divergent or noncontinuous interaction potentials. 
Moreover, it  enables analytical investigations as discussed in \red{Section}~\ref{sec:discussion}. 
Note that the interaction energy is structurally comparable to that used to describe soft deformable particles via a Gaussian interaction potential in~\cite{menzel2012soft,tarama_individual_2014}, however, the resulting expressions differ in details~\footnote{To understand in detail the  differences, we refer the reader to the respective literature: Eq.~(4) in~\cite{menzel2012soft}, Eqs.~(20)-(24) in~\cite{tarama_individual_2014} and Eqs.~(7)-(11) in~\cite{ohta2014traveling} should be compared to the equations given in the Supplemental Material of this work~\cite{SI}.}.

\paragraph*{Equations of motion}

We describe the dynamics of spatially extended, active particles in the overdamped limit as follows~\cite{doi_theory_1986,peruani_nonequilibrium_2006,baskaran_hydrodynamics_2008,abkenar_collective_2013,weitz_selfpropelled_2015,shi_self_2018}: 
\begin{subequations}
\label{eqn:mot}
\begin{align}
	\dot{\vec{r}}_k &= v_0 \vec{e}_{\!\!\:\para} \! \left [ \varphi_k \right ] + \hat{\boldsymbol{\mu}} \! \left[ \varphi_k \right] \!\!\: \cdot \!\!\: \vec{F}_{k} + \!\: \boldsymbol{\eta}_k(t), \label{eqn:spat_dyna} \\
	\dot{\varphi}_k &= \mu_{\varphi} M_k + \! \sqrt{2 D_{\varphi} } \!\; \xi_k \! \left( t \right) \! .\label{eqn:ang_dyna}
\end{align}
\end{subequations}
Accordingly, the balance of dissipative and driving force leads to stochastic motion with a mean speed~$v_0$ in the absence of interactions. 
The translational ($\hat{\boldsymbol{\mu}}$) and rotational~($\mu_{\varphi}$) mobilities are determined by the anisotropy of particles as well as the properties of the surrounding medium; the particular functional forms depend on the experimental context~\cite{perrin_mouvement_1934,*koenig_brownian_1975,tirado_comparison_1984,doi_theory_1986}. 
Similarly, the nature of fluctuations of position and orientation, abbreviated by~$\boldsymbol{\eta}_k(t)$ and $\xi_k \! \left( t \right)$, respectively, depends on the context.
Here, they are assumed to be Gaussian, unbiased and $\delta$-correlated in time.
Furthermore, fluctuations of the center of mass are anisotropic,
\begin{align}
	\label{eqn:def:noise_corr}
 \boldsymbol{\eta}_k(t) \!\!\: = \!\!\: 
     \sqrt{2D_{\para}} \!\: \vec{e}_{\para} \!\!\: \! \left [ \varphi_k \right ] \!\!\: \eta_{\para,k}(t) 
   \!\!\:+\!\!\: \sqrt{2D_{\perp}} \!\: \vec{e}_{\perp} \!\!\: \! \left [ \varphi_k \right ] \!\!\: \eta_{\perp,k}(t). 
\end{align}
where the diffusion coefficients parallel and perpendicular with respect to the rods' orientation read $D_{\para}$ and $D_{\perp}$. 

\begin{figure*}[t]
 	\begin{center}
 		\includegraphics[width=\textwidth]{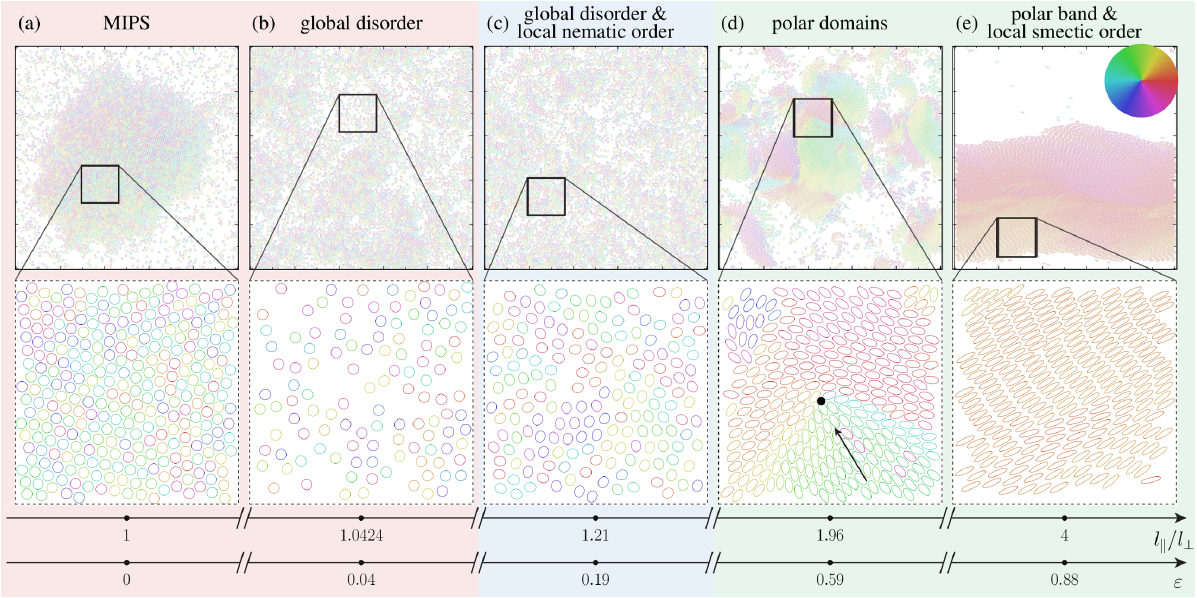}
 	\end{center}
	\vspace{-0.5cm}
	\caption{Snapshots of large-scale patterns for increasing particle anisotropy of rigid self-propelled objects. Bottom row shows  enlarged~($40 \times 40$) images. The anisotropy is varied keeping the particle size $A \propto l_{\parallel} l_{\perp}$ fixed by setting the product $l_{\parallel} l_{\perp} = 1$. From left to right: $\varepsilon = 0$, aspect ratio $l_{\parallel} / l_{\perp} = 1$; $\varepsilon \approx 0.04$, aspect ratio $l_{\parallel} / l_{\perp} = 1.0424$; $\varepsilon \approx 0.19$, aspect ratio $l_{\parallel} / l_{\perp} = 1.21$; $\varepsilon \approx 0.59$, aspect ratio $l_{\parallel} / l_{\perp} = 1.96$; $\varepsilon \approx 0.88$, aspect ratio $l_{\parallel} / l_{\perp} = 4$. Other parameters: energy functional $\mathcal{F} \! \left [ \xi \right ] = \xi^{\gamma}$ with energy scale $\kappa = 1$ and exponent $\gamma = 3$, active force $F_a = v_0 / \mu_{\parallel} = 0.01$, spatial diffusion $D_{\parallel,\perp} = 0$, rotational fluctuations $D_{\varphi} = 3 \!\!\;\cdot \!\!\; 10^{-5}\mu_{\varphi}$, systems size $L_{x,y} = 250$ and particle number $N = 5968$. The mobility matrix for ellipsoids dispersed in a liquid was used~\cite{oberbeck_ueber_1876,edwardes_steady_1892,perrin_mouvement_1934,*koenig_brownian_1975}. Color coding represents the orientation of the active force. See \red{Appendix}~\ref{app:trans-pol-ord} for additional snapshots showing the gradual emergence of polar order. The background color indicates the different physical regimes identified in Fig.~\ref{fig:PhaseDiagram}. For movies, see~\cite{SI}.}
	\label{fig:mips_rods}
\end{figure*}

We point out that interaction as well as the diffusion matrix and the friction tensor possess ``nematic symmetry''~--~they are invariant under the transformation~$\varphi_j \rightarrow \varphi_j + \pi$ for any rod. 
Notably, the self-propulsion term $v_0 \vec{e}_{\!\!\:\para} \! \left [ \varphi_j \right ]$ in Eq.~\eqref{eqn:mot} breaks this inversion symmetry at the microscale as~$\vec{e}_{\!\!\:\para} \! \left [ \varphi_j \right ] \rightarrow \vec{e}_{\!\!\:\para} \! \left [ \varphi_j + \pi \right ] = - \vec{e}_{\!\!\:\para} \! \left [ \varphi_j \right ]$. 
For this reason, self-propelled rods ($v_0 > 0$) are inherently different from  systems without directed self-propulsion on the microscale ($v_0 = 0$), where the diffusive dynamics is invariant under inversions of the orientation vector; this applies to most active nematic models~\cite{narayan_long_2007,giomi_excitable_2011,giomi_defect_2013,decamp_orientational_2015,shi_topological_2013,thampi_instabilities_2014,putzig_instabilities_2016,oza_antipolar_2015,pismen_viscous_2017,cortese_pair_2018}.

\section{Unifying active matter}
\label{sec:unfy_pat}

The smoothed-particle approach allows to unify several seemingly disconnected paradigmatic phenomena reported for different systems before within one framework, such as active phase separation, polarly moving clusters, nematic bands as well as the coexistence of polar and nematic order.
%
%
Generally, the emergent patterns depend on particle shape and are sensitive to whether particles can slip over each other~--~implying that the self-propulsion force can overcome the repulsive interaction force~--~or not~\cite{shi_self_2018}.  
The ratio between the strength of the active force and the maximum repulsion force in a binary collision defines the \textit{rigidity} of the particles. 
We report the phenomenology of this system in the regime of weak self-propulsion, i.e.~high repulsion force implying high rigidity, for different particle shapes in \red{Section}~\ref{sec:pheno} and turn the role played by self-propulsion in \red{Section}~\ref{sec:nema_polar}.

\subsection{From self-propelled discs to self-propelled rods}
\label{sec:pheno}

%
The shape of rigid self-propelled objects determines their collective behavior.
%
%
%
Below, we describe the phenomenology observed in numerical simulations along Fig.~\ref{fig:mips_rods} as the anisotropy of rigid particles is increased. 
%

\paragraph*{Breakdown of active phase-separation}

\begin{figure*}[t]
 	\begin{center}
		\includegraphics[width=\textwidth]{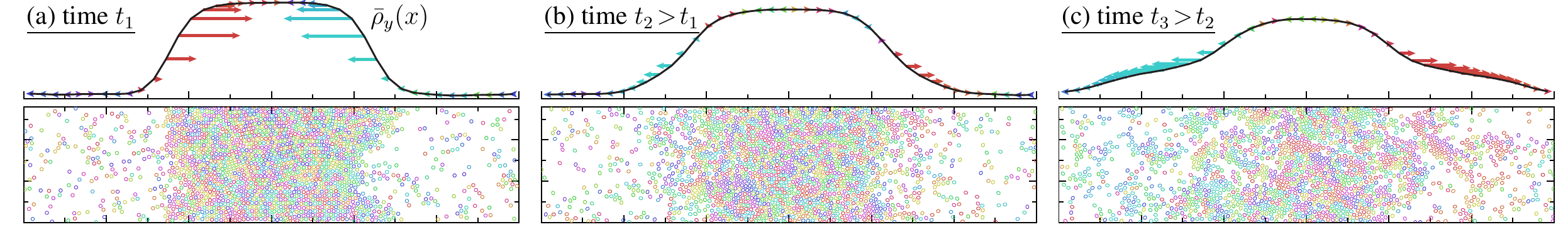}
 	\end{center}
	\vspace{-0.65cm}
	\caption{Unstable MIPS aggregate after a quench to slightly anisotropic rods~($l_{\parallel} = 1.1, l_{\perp} = 0.9$):~a cross section of the density field and the $p_x$-component of the polar order parameter field in a slap geometry and corresponding snapshots  for three different times~$t_1 < t_2 < t_3$. The state on the right is not a stationary state which is given by a flat density profile. Parameters:~energy functional $\mathcal{F} \! \left [ \xi \right ] = \xi^{\gamma}$ with energy scale $\kappa = 1$ and exponent $\gamma = 3$, active force $F_a = v_0 / \mu_{\parallel} = 0.01$, particle diameter $l_{\parallel} = l_{\perp} = 1$, aspect ratio $l_{\parallel} / l_{\perp} = 1$, anisotropy $\varepsilon =0$, spatial diffusion $D_{\parallel,\perp} = 0$, rotational diffusion $D_{\varphi} = 7.5 \cdot 10^{-5}$, systems size $L_{x} = 300$ and $L_y = 70$, particle number $N = 1671$. }
	\label{fig:mips_slap_stab}
\end{figure*}

In the limiting case of self-propelled discs~($l_{\para} / l_{\perp} = 1$), we observe phase separation due to active motion into a disordered, gas-like phase and hexatically ordered aggregates, called motility-induced phase separation~(MIPS)~\cite{tailleur_statistical_2008,cates_mips_2015,nardini_entropy_2017,solon_generalized_2018,solon_generalized2_2018,prymidis2016,prymidis2017,bialke_microscopic_2013,haertel_three_2018,haertel_three_2018}.
In Fig.~\ref{fig:mips_rods}\red{(a)}, the enhancement of density fluctuations in the fully phase-separated regime is evident. 
Since the orientations of discs within an aggregate is disordered, this phenomenon may therefore be described by a scalar field theory for the particle density only~\cite{solon_generalized2_2018}.

%
Surprisingly, MIPS aggregates are found to melt for small anisotropies of the active particles.
In contrast to the phase-separated regime, we observe a drastic decrease in density fluctuations as well as a reduction of local hexatic order already for weakly anisotropic particles~(Fig.~\ref{fig:mips_rods}\red{(b)}, where~$l_{\para} / l_{\perp} = 1.0424$).
See also \red{Appendix}~\ref{app:trans-pol-ord} for additional information on  the hexatic order.

What is the mechanism behind break up of aggregates? 
We recall that phase-separation of self-driven spheres arises due to the slow down of particles as they collide.
Upon a collision, their orientations point towards the center of clusters such that aggregates are surrounded by a polar boundary layer on average, whereby an active pressure keeps them together~[Fig.~\ref{fig:mips_slap_stab}\red{(a)}]. 
Preparing an aggregate in a slap geometry and performing a quench to slightly anisotropic rods~[Figs.~\ref{fig:mips_slap_stab}\red{(b,c)}] reveals that this polar boundary layer dissolves as a deterministic torque rotates rods away from the boundary of aggregates~[Fig.~\ref{fig:mips_slap_stab}\red{(b,c)}].

\paragraph*{Emergence of orientational order}

We observe the formation of states with orientational order by increasing the aspect ratio of particles beyond the breakdown of active phase separation.  
At first, the system remains globally disordered, however, it displays weak local nematic order as revealed by the semi-analytical analysis discussed in \red{Section}~\ref{sec:em_ord}. 
Counterintuitively, local order becomes polar if the anisotropy is increased further, even though the interaction is strictly nematic. 
The emergence of local polar order is illustrated in Fig.~\ref{fig:mips_rods}\red{(c-e)} as well as in \red{Appendix}~\ref{app:trans-pol-ord} for intermediate aspect ratios.
In particular, polarly ordered, moving domains are found~\cite{peruani_nonequilibrium_2006}.
Those macroscopic patterns, as shown in Fig.~\ref{fig:mips_rods}\red{(d)}, are highly dynamic since polar order is inherently related to mass transport, thereby inducing clusters to form, merge and break in a nontrivial fashion~\cite{peruani_kinetic_2013}.

Along with polar domains, topological defects emerge due to collisions of those structures  as highlighted by a black dot in Fig.~\ref{fig:mips_rods}\red{(d)}.
Examining the orientation of the rod axis only, i.e.~irrespective of the orientation of the self-propulsion force with respect to the body axis, these defects have basically a nematic structure.
Defects may, however, be self-motile because of the polarity of directional energy input at the level of individual rods:~in Fig.~\ref{fig:mips_rods}\red{(d)}, a black arrow indicates that rods push towards the center of a $+\frac{1}{2}$-defect thereby creating an active, anisotropic stress, which results in a directed displacement of the defect position; an analogous defect dynamics was observed in~\cite{genkin_topological_2017}.
This mechanism of defect motion in ensembles of polarly driven objects is different from defect motility in \textit{active nematics}, both dry~\cite{narayan_long_2007} and wet~\cite{giomi_defect_2013,thampi_instabilities_2014,pismen_viscous_2017,cortese_pair_2018}. 
Furthermore, defects are created and disappear in an intermittent way: due to strong density instabilities, the defects may vanish in the void or penetrate from the boundary of a dense region~--~the topological charge is therefore not conserved.

Polar domains may become system spanning for intermediate system sizes as shown on the right of Fig.~\ref{fig:mips_rods} for $l_{\para} / l_{\perp} = 4$.
These polar bands are comprised of smectic particle arrangements~\cite{chaikin_principles_2000,weitz_selfpropelled_2015,romanczuk_smectic_2016}. 
Note, however, that numerical data suggest the absence of long-range orientational order in the thermodynamic limit~\cite{weitz_selfpropelled_2015}.

Let us stress that the observed macroscopic order is polar, while the symmetry of interaction potential  is strictly nematic~[cf.~Eq.~\eqref{eqn:mot}].
This implies that the symmetry of the macroscopic order is not imposed by the symmetry of the interaction potential, but emerges spontaneously from the spatial dynamics. 
Similar behavior was observed experimentally in studies of colonies of myxobacteria as well as corresponding hydrodynamic descriptions~\cite{peruani_collective_2012,harvey2013continuum}.

\subsection{Polar vs.~nematic order and their coexistence}
\label{sec:nema_polar}

\begin{figure*}[t]
 	\begin{center}
 	\includegraphics[width=\textwidth]{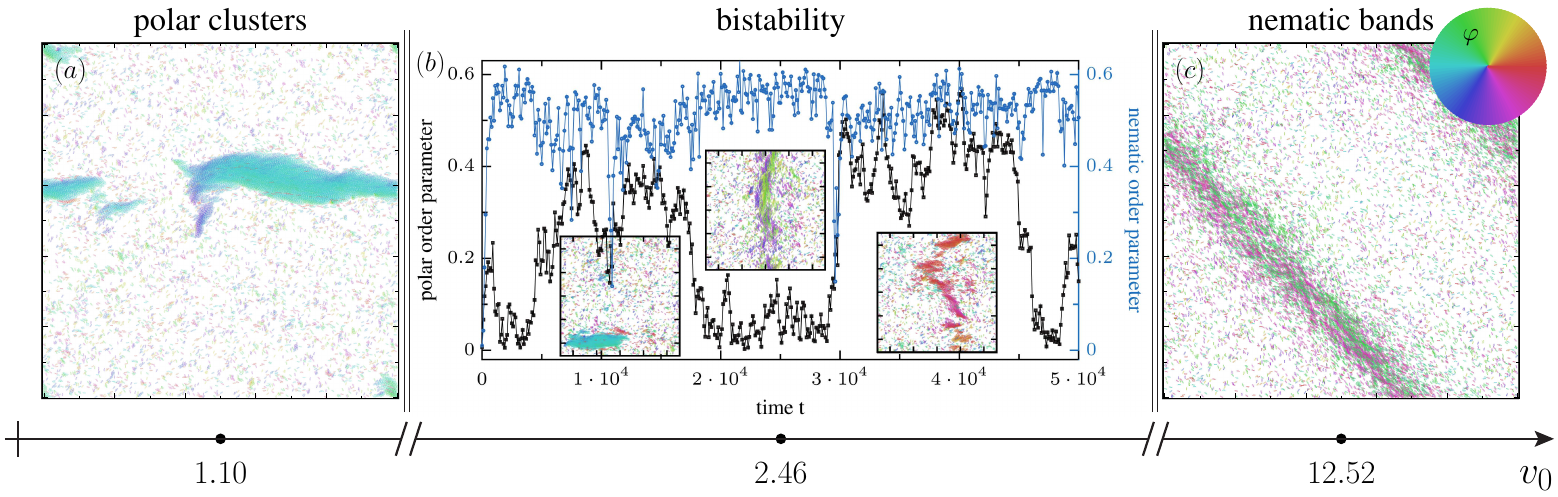}
 	\end{center}
	\vspace{-0.5cm}
	\caption{From polar to nematic order via their coexistence:~snapshots of large-scale patterns observed in numerical simulations of rods as a function of their self-propulsion speed for fixed repulsion strength and anisotropy. If the self-propulsion is small, such that repulsion forces cannot be overcome~(left), polarly ordered domains are observed. In contrast, nematic bands~--~previously reported for Vicsek-type particles with nematic velocity alignment~\cite{ginelli_large_2010}~--~emerge as the high self-propulsion force allows particles to glide over each other. Surprisingly, we find a bistable coexistence of nematic bands and polar clusters for intermediate values of the self-propulsion force: the nematic order parameter~$\abs{\sum_{j=1}^N e^{2 i \varphi_j} \!\!\: / N}$ fluctuates around a constant value, whereas the polar order parameter~$\abs{\sum_{j=1}^N e^{i \varphi_j} \!\!\: / N}$ switches stochastically between two values, corresponding to one state with polar order~(nonzero value of polar order parameter) to a nematic state where the polar order parameter fluctuates around zero. The respective snapshots are shown as insets. Parameters:~fixed particle shape~$A \propto l_{\parallel} l_{\perp}$ by~$l_{\parallel} l_{\perp} = 1$, anisotropy $\varepsilon \approx 0.88$, aspect ratio~$l_{\parallel} / l_{\perp} = 4$, energy functional~$\mathcal{F} \! [\xi] = \xi^\gamma$ with energy scale~$\kappa = 1$ and $\gamma = 3$, spatial diffusion~$D_{\parallel,\perp} = 0$, rotational diffusion~$D_{\varphi} = 0.022$, system size~$L_{x,y} = 500$ in (a) and (c), $L_{x,y} = 250$ in (b), particle density~$\rho_0 = 0.08$. } 
	\label{fig:NemaToPolarOrder}
\end{figure*}

Let us now fix the aspect ratio of particles and ask what the influence of the rigidity of particles is.
In order to answer this question, we varied the self-propulsion force resulting in higher speeds $v_0$.
In the limit of high self-propulsion, particles may slide over each other upon encounter~--~an effect reminiscent of tunneling in quantum mechanics which is possible due to the particle-field representation of soft rods~--~whereas they would be blocked by their interaction partners in the opposite limit. 
Fig.~\ref{fig:NemaToPolarOrder} shows pictorially the phenomenological transition from rigid to soft rods.
Large-scale, polar domains are observed for low self-propulsion. 
In the limit of high activity, in contrast, particles arrange themselves in nematic band-like structures as they are familiar from Vicsek-type, point-like particles with nematic alignment~\cite{ginelli_large_2010}. 
Surprisingly, we found a bistable coexistence region if the order of magnitude of the self-propulsion force is comparable to repulsive forces:~nematic bands and polarly ordered domains are observed in an intermittent fashion.

The coexistence of polar and nematic order is detailed in Fig.~\ref{fig:NemaToPolarOrder}\red{(b)}. 
The stochastic switching from polar to nematic states is revealed by anomalously high fluctuations of the polar order parameter. 
The timescales of these stochastic transitions are remarkable, as they are several orders of magnitude larger than microscopic timescales of the dynamics at the particle level.

The simultaneous existence of polar and nematic states has recently been reported by Huber et al.~in~\cite{huber_emergence_2018} for a motility assay experiment.
Those results were rationalized by simulations of self-propelled, flexible filaments which are pulled at one side and interact by volume exclusion.
In contrast to this system, the interaction of the self-propelled particles considered in the present study is strictly nematic.
The global nematic symmetry is solely broken by the self-propulsion term; individual entities are nematic in contrast to the filaments in~\cite{huber_emergence_2018}.
Here, we show for the first time that bistability of polar and nematic structures can also be expected for simple self-propelled rods if the strength of self-propulsion and repulsion are fine-tuned or happens to coincide in a specific application.

We conclude this section by a comment on finite size scaling:~we do not expect the bistability of globally polar and nematic states to be retained in the thermodynamic limit as the diffusive motion of particles within the band is too slow to allow for long-range nematic order~\cite{weitz_selfpropelled_2015}.
We do rather expect disconnected patches composed of polarly or nematically aligned particles to emerge which may coarsen in a nontrivial way as they interact at the mesoscale. 
Due to the enormous length-scale separation of several orders of magnitude, a detailed analysis of this phenomenon is beyond the scope of the present work.

\section{Analytical insights} 
\label{sec:discussion}

The theoretical framework based on a smoothed-particle representation allows for a semi-analytical analysis within kinetic and hydrodynamic theories derived from the microscopic dynamics. 
In the following, we elucidate the large-scale pattern formation in the course of the transition from active spheres to self-propelled rods in two subsequent steps:~the breakdown of MIPS due to combined action of self-propulsion and anisotropic particle shape and the emergence of orientational order from volume exclusion interactions.
Afterwards, we conclude by discussing the intricate relation of symmetry of the particle-particle interaction and the symmetries of emergent patterns at large scales in active matter.

The theory is based on the nonlinear Fokker-Planck equation for the one-particle density~$P \! \left( \vec{r},\varphi,t \right)$,
\begin{align}
	\label{eqn:FP1}
	\partial_t P \!\:\!=\!\:\! &- \!\!\: \nabla \!\!\:\cdot\!\!\: \Big [ \big ( v_{0}\vec{e}[\varphi] + \hat{\boldsymbol{\mu}} \! \left [ \varphi \right ] \!\!\: \cdot \!\!\: \vec{F} \big ) P \Big ] \! 
	+  \! \nabla \!\cdot \! \Big [\mathcal{D} \! \left [ \varphi \right ] \!\!\:\cdot\! \nabla P \Big ] \\
	&-\!  \partial_{\varphi} \big [ \mu_{\varphi} M P \Big ] 
	+ D_{\varphi} \partial_{\varphi} P, \nonumber
\end{align}
with the effective force and torque functionals
%
\begin{align*}
	\vec{F} &= \! \int \!\!d^{\!\:2} \!\!\; r' d\varphi' \, \vec{f}_{2} \! \left( \vec{r} - \vec{r}' \!,\varphi,\varphi' \right) \!\!\: P\!\left( \vec{r}' \!,\varphi'\!,t \right) g_2(\vec{r},\vec{r}'\!;\varphi,\varphi'\!,t), \\
	M &= \! \int \!\!d^{\!\:2} \!\!\; r' d\varphi' \, m_{2} \! \left( \vec{r}-\vec{r}'\!,\varphi,\varphi' \right) \!\!\: P\!\left( \vec{r}' \!,\varphi'\!,t \right) g_2(\vec{r},\vec{r}'\!;\varphi,\varphi'\!,t),
\end{align*}
%
which, in turn, depend on pair distribution function~$g_2$.
We point out at this level that the binary interaction $\vec{f}_2$ and $m_2$, i.e.~the force and torque which one particle exerts on another interaction partner~[cf.~Eq.~\eqref{eqn:bin_torques}], are effectively renormalized by the emergent correlations quantified by~$g_2$.
Hence, the interplay of self-propulsion and volume exclusion interactions as well as the collision kinetics can induce novel terms in mesoscale, coarse-grained descriptions, which are not of nematic symmetry, whereas the interaction at the particle level is strictly nematic.

\subsubsection{Breakdown of active phase separation}
\label{par:sph_lim}

At first, we study the breakdown of motility-induced phase separation~(MIPS) for small particle anisotropies as illustrated in Figs.~\ref{fig:mips_rods}, \ref{fig:mips_slap_stab}.
For the sake of simplicity, we only consider leading orders in~$\varepsilon$ (\textit{weakly non-spherical limit}).
This approximation is intended to explain why MIPS aggregates in the classical sense are unstable for anisotropic particles.

In the spherical limit, the force at leading order reduces to an isotropic central body force.
Notably, there is a qualitatively novel contribution to the torque at leading order~$\varepsilon$ which has not been studied analytically so far in the context of active matter to the best of our knowledge:~the torque felt by particle~$k$ due to the presence of particle~$j$ is proportional to
\begin{align}
	m_{j \rightarrow k} \propto \varepsilon \sin \! \Big [ 2 \big( \varphi_k \!\!\: - \arg \! \left[ \vec{r}_k - \vec{r}_j \right ] \!\!\: \big ) \!\!\: \Big ] ,
\end{align}
cf.~Eq.~\eqref{eqn:bin_torques}. 
Its meaning and interpretation as \textit{collision avoidance} is the central subject of the discussion below. 
Note that the nematic alignment is not relevant in the spherical limit as it is of higher order in~$\varepsilon$. 
Furthermore, mobility and diffusion tensor of individual particles are assumed to be approximately isotropic:~$\hat{\boldsymbol\mu} \approx \bar{\mu} \mathds{1}$ and~$D_0 = D_{\para} \approx D_{\perp}$.

We begin the analysis by calculating the average force and torque~--~felt by a particle with orientation~$\varphi$~--~ to first order in gradients:  
\begin{subequations}
\label{eqn:sl_ft}
\begin{align}
\label{eqn:sl_ft_a}
 	\!\!\!\vec{F} &\simeq - \zeta_0 \kappa \vec{e}_{\!\!\:\para} \! \left [ \varphi \right ] \! \rho, \! \\
 	\!\!\!M       &\simeq - \varepsilon \zeta_1 \kappa \vec{e}_{\!\!\:\para} \! \left [ \varphi \right ] \!\!\:\wedge \!\!\: \!\!\: \nabla \rho \!\!\:=\!\!\: - \varepsilon \zeta_1 \kappa \! \left( \cos \varphi \partial_y \!\!\: - \sin \varphi \partial_x \right) \!\!\: \rho. \! \!\!
	\label{eqn:sl_ft_b}
\end{align}
\end{subequations}
The parameters~$\zeta_{0,1}$ are nonequilibrium transport coefficients which can be expressed as integrals over the inter-particle correlation function and~$\rho$ is the particle density; for details on the derivation, see~\cite{SI}.
Numerical measurements of pair-correlations within the particle-based dynamics~(see Fig.~\ref{fig:g2_spheres}) reveal that the kinetic of particles is such that leads to an enhancement of particle density at the front with respect to the direction of motion of a focal particle: as particles move actively in a semi-dilute environment, they tend to collide with others, and consequently the probability to find a particle in front is higher than in the back.  
This implies positive transport coefficients~$\zeta_{0,1}$. 
This phenomenon, i.e.~accumulation of particle density at the front of the moving particle, was observed in system of self-propelled discs and used to build a scalar field theory to describe phase separation in active disk systems~\cite{bialke_microscopic_2013,haertel_three_2018}.  
According to Eq.~\eqref{eqn:sl_ft_a}, the force yields, to first order, an average slowing down of particle speed in high density areas as they bump into their neighbors.
The decrease of speed with particle density is the classical mechanism underlying MIPS, first introduced in~\cite{tailleur_statistical_2008}. 
In Eq. \eqref{eqn:sl_ft_b}, we report, however, an important new element not present in classical theories for MIPS:~a slight asymmetry of the body shape gives rise to a torque, which induces a rotation away from high density domains.

The physical effects of force and torque become evident at the level of coarse-grained order parameters:~the density $\rho \! \left( \vec{r},t \right)$ and the polar order parameter field~$\vec{p} = \mean{\vec{e}_{\!\!\:\para} \! \left [ \varphi \right ]}$. 
Equations for the time evolution of these quantities are obtained by performing a mode expansion of the Fokker-Planck equation~\eqref{eqn:FP1} which yields
\begin{subequations}
\label{eqn:hydro:sph}
\begin{align}
	\partial_t \rho & \!\!\:\approx\!\!\: - \nabla \cdot \big [ v(\rho) \vec{p} \big ] + \!\!\: D_0 \Delta \rho , \label{eqn:mode_r} \\ 
	\partial_t \vec{p} & \!\!\:\approx\!\!\: - \nabla \!\!\: \cdot \!\!\: \left [ \frac{v(\rho)}{2} \, \Pi_{+} \right ] \! - \! D_{\varphi} \vec{p} \!\!\:-\!\!\: \frac{\mu_{\varphi}\varepsilon \zeta_1 \kappa}{2}\, \Pi_{-} \! \cdot \!\!\: \nabla\!\!\: \rho + \!\!\: D_0\Delta \vec{p}.\!\! \label{eqn:mode_p}
\end{align}
\end{subequations}
The convective term in Eq.~\eqref{eqn:mode_r} for the  density represents  the density-dependent speed reduction due to collisions via $v \! \left( \rho \right) = v_0 - \bar{\mu} \zeta_0 \kappa \rho$.
A corresponding term is found in Eq.~\eqref{eqn:mode_p} for the polar order parameter as well. 
We find that the torque is cast at the field level as an anisotropic, nonlinear flow of the form $\dot{\vec{p}} \propto - \varepsilon \Pi_{-} \! \cdot \!\!\;\nabla \rho$ with the tensors $\Pi_{\pm} = \rho \mathds{1} \pm \mathfrak{Q}$, where~$\mathfrak{Q} \! \left( \vec{r},t \right)$ abbreviates the nematic order parameter field.
Accordingly, the torque is such that the density gradients are counteracted by an opposing particle flow.
We stress that the polar and nematic order parameter have to be taken into account for  the anisotropic, nonequilibrium stresses due to active motion on the hydrodynamic level.

\textit{Robustness of MIPS} 
The time-independent solutions of the transport equations~\eqref{eqn:hydro:sph} imply the polar order parameter~$\vec{p}$ to be collinear to density gradients~$\nabla \rho$, as~$\vec{p} = D_0 \!\left (\nabla \rho \right ) \! / v \! \left( \rho \right)$ follows from Eq.~\eqref{eqn:mode_r}, cf.~the phase-separated state in Fig.~\ref{fig:mips_slap_stab}\red{(a)}. 
The theoretical analysis for anisotropic particles reveals, however, that torques will destabilize parallel arrangements of the orientation $\vec{e}_{\!\!\:\para} \! \left [ \varphi \right ]$ and density gradients $\nabla \rho$ [see Eq.~\eqref{eqn:sl_ft_b}].
Therefore, this torque, which is proportional to the anisotropy~$\varepsilon$, tends to dissolve the polar boundary layer around aggregates.
Namely, it induces locally anisotropic stresses whenever density gradients and local order coexist on a coarse-grained level~[cf.~Eq.~\eqref{eqn:mode_p}], as we argued before based on numerical simulations (see \red{Section}~\ref{sec:pheno}).

To substantiate these arguments, the linear stability of the spatially homogeneous, disordered state is investigated on the basis of Eqs.~\eqref{eqn:hydro:sph}.
For self-propelled spheres, the emergence of MIPS is signaled by a long-wavelength instability of the density field~\cite{bialke_microscopic_2013}.
Here, we examine  the stability of the disordered state with respect to long-wavelength fluctuations for anisotropic particles.
For this purpose, the dynamics of the polar order parameter field is linearized first by inserting $\rho = \rho_0 + \delta \rho$ and $\vec{p} = \delta \vec{p}$. 
As we are interested in the onset of a long-wavelength instability, the linearized field~$\delta \vec{p}$ can further be adiabatically eliminated yielding
\begin{align}
	\label{eqn:adiael:dp}
	\delta \vec{p} \simeq - \frac{(v_0 - 2\bar{\mu}\zeta_0 \kappa \rho_0) + \varepsilon \mu_{\varphi} \zeta_1 \kappa \rho_0}{2 D_{\varphi}} \!\: \nabla \delta \rho. 
\end{align}
In order to determine the onset the long-wavelength instability, it is not necessary to take the fluctuations of the nematic order parameter~$\mathfrak{Q}$ into account, which was therefore set to zero. 
To leading order, one thus obtains an effective diffusion equation~$\partial_t \delta \rho \simeq \Gamma \Delta \delta \rho$ for the fluctuations of the density around the spatially homogeneous state by inserting  this expression into Eq.~\eqref{eqn:mode_r}, where the transport coefficient~$\Gamma$ reads
\begin{align}
	\Gamma = D_0 + \frac{ \left( v_0 - \bar{\mu} \zeta_0 \kappa \rho_0 \right) \! \left [ \left( v_0 - 2 \bar{\mu} \zeta_0 \kappa \rho_0 \right) \! + \varepsilon \mu_{\varphi} \zeta_1 \kappa \rho_0 \right ]}{2 D_\varphi}. 
\end{align}
A long-wavelength instability of the homogeneous state towards a phase-separated state occurs for~$\Gamma < 0$; the spinodal curves are determined by~$\Gamma = 0$. 
Since this instability condition depends explicitly on $\varepsilon$, we can derive a critical particle anisotropy above which MIPS cannot arise~(sufficient condition):  
\begin{align}
	\label{eqn:exist_crit}
	\varepsilon  > \frac{v_0 - 4 \sqrt{D_0 D_{\varphi}}} {\mu_{\varphi} \rho_0 \zeta_1 \kappa} \,.
\end{align}
Thus, there is a critical anisotropy beyond which the polar boundary layer which would keep an aggregate together becomes destabilized by torques.
In agreement with these theoretical arguments, we find in simulations of the model that MIPS aggregates do not emerge already for rather small anisotropies~--~for the parameters used in this study, we find the critical values $\epsilon \gtrsim 0.04$, correspondingly $l_{\parallel}/l_{\perp} \gtrsim 1.04$, see Fig.~\ref{fig:mips_rods}\red{(b)}; see also the dissolution of a MIPS aggregate after a quench from spherical particles to anisotropic rods (Fig.~\ref{fig:mips_slap_stab}).
These findings put the relevance of the classical phenomenon of MIPS for self-driven, anisotropic particles, such as self-propelled rods, into question.

\subsubsection{Onset of orientational of order}
\label{sec:em_ord}

Let us now examine the  emergence of orientational order beyond the spherical limit, as observed numerically for large anisotropies (see Fig.~\ref{fig:mips_rods}). 
As the interaction at the particle level possesses nematic~(front-tail) symmetry~[Eq.~\eqref{eqn:epot_u2}], one may naively expect the emergence of local nematic order.
We will show that the break up of MIPS is indeed followed by a globally disordered phase with local nematic order. 
Interestingly, local order becomes, counter-intuitively, polar if the aspect ratio is increased even further. 
In order to identify and understand the emergence of local orientational order at the hydrodynamic level, we derived coarse-grained order parameter equations  where hydrodynamic transport coefficients are expressed as integrals over the correlation functions. 
To simplify the presentation, we concentrate on central, symmetry-breaking terms for the polar and nematic order at the local level, i.e.~we expand to lowest order in spatial gradients:
\begin{subequations}
	\label{eqn:hydroP}
	\begin{align}
	 \dot{\vec{p}}        &= \sigma_p \vec{p} + \mathcal{O} \! \left( \nabla \right) \! , \\
	 \dot{\mathfrak{Q}}   &= \sigma_n \mathfrak{Q} + \mathcal{O} \! \left( \nabla \right) \! .
	\end{align} 
\end{subequations}
If the transport coefficient~$\sigma_p$ is positive, the local polar order parameter grows and, thus,  ordered polar structures are  expected at local scales.
In contrast, the nematic order parameter is relevant at the local level if $\sigma_p <0 $ and $\sigma_n > 0$.

The starting point of the analysis is the Fokker-Planck equation~[Eq.~\eqref{eqn:FP1}] for the one-particle density. 
A mode expansion of this Fokker-Planck equation, where only the relevant terms under consideration are kept, yields the following expressions for the transport coefficients: 
\begin{subequations}
	\label{eqn:orientation:tc}
\begin{align}
	\sigma_p & \!=\!\!\: \frac{\rho_0 \mu_{\varphi}}{2\pi} \!\int_{0}^{\infty} \!\!\!dr \, r \!\!\!\: \int_{0}^{2\pi} \!\!\! \!\!\: d\alpha  \!\!\: \! \int_{0}^{2\pi} \!\!\! \!\!\: d\varphi \, \sin \! \left( \varphi \right) \tilde{m}_2 \! \left( r,\alpha,\varphi \right) \!\!\: - \!\!\: D_{\varphi} , \label{eqn:orientation:tca} \\
	\sigma_n & \!=\!\!\: \frac{\rho_0 \mu_{\varphi}}{\pi} \!\int_{0}^{\infty} \!\!\!dr \, r \!\!\!\: \int_{0}^{2\pi} \!\!\! \!\!\: d\alpha \!\!\: \! \int_{0}^{2\pi} \!\!\! \!\!\: d\varphi \, \sin \! \left( 2\varphi \right) \tilde{m}_2 \! \left( r,\alpha,\varphi \right) \!\!\: - \!\!\: 4 D_{\varphi} .  \label{eqn:orientation:tcb}
\end{align}
\end{subequations}
In these equations, we introduced $\tilde{m}_2 = m_2 g_2$ which is the product of the actual torque~$m_2$ between two particles and the pair distribution function~$g_2$.
It can be interpreted as an \textit{effective mean-field model}.
We implicitly assumed that the pair-distribution function~$g_2$ is known and absorbed it into the definition of~$\tilde{m}_2$. 
Thereby, the transport coefficients~$\sigma_{p,n}$ above still depend on the inter-particle correlations and the collision kinetics.

The simplest mean-field approximation assumes that $g_2 \approx 1$~--~consequently, the effective torque~$\tilde{m}_2$ is identical to the actual torque~$m_2$.
In this limit, the transport coefficient~$\sigma_p$ for the polar order parameter is always negative; the integral in Eq.~\eqref{eqn:orientation:tca} vanishes for symmetry reasons:~$\sigma_p = -D_{\varphi}$. 
The emergence of polar order cannot be described within mean-field theory; it only predicts the existence of an isotropic-nematic transition~\cite{baskaran_enhanced_2008,baskaran_hydrodynamics_2008}.

How can one rationalize the emergence of polar order provided that the interaction among two anisotropic, elongated particles is strictly nematic?
We first recall that only the self-propulsion force breaks the nematic symmetry of the microscopic dynamics~[Eq.~\eqref{eqn:mot}].
Accordingly, the emergence of polar order has to be related to the collision kinetics and, in particular, the formation of polar clusters~\cite{peruani_nonequilibrium_2006} which is, in turn, reflected by correlations which need to be taken into account properly.

Along with the illustration of polar and anti-polar collisions in Fig.~\ref{fig:RodEx}, we first give a heuristic argument for the emergence of local polar order in ensembles of rigid rods.
Let us consider a collision under an acute angle:~an elongated, active particle that collides with a cluster aligns its direction of motion to the local mean orientation.
Consequently, it will keep on moving in parallel with this cluster for a significant time.
Only rotational diffusion  may deflect  its direction of motion away from the boundary of the cluster.
Therefore, polar clusters are stable and may grow for low rotational noise.
In contrast, nematic clusters cannot exist:~a particle that collides in an antipolar way with a cluster will just slide off its boundary due to the mobility of the cluster.
As the distance of the particle and the cluster increases over time, their positions decorrelate as a result of collisions with other particles. 
This reasoning leads to the conclusion that the probability to find rigid rods moving in parallel is higher than seeing antipolar arrangements locally as a result of the collision kinetics.
This can be verified quantitatively by measuring the pair correlation function numerically, see supplementary Fig.~\ref{fig:g2_spheres}\red{(b)}.

More formally, the effective, binary torque on the field level~$\tilde{m}_2$, which enters into the relevant hydrodynamic transport coefficient~$\sigma_p$, is given by the product of the model~$m_{2}$ and the correlation function~$g_2$.
In turn,  this quantity encodes the kinetics of collisions.
Consequently, the effective torque $\tilde{m}_2$ may contain new terms which are not present on the particle level:~the torque at the particle level contains nematic alignment as~$m_{2} \! \propto \sin \left [ 2 \! \left( \varphi' - \varphi \right) \right ]$, and the pair distribution function contains a positive contribution~$g_2 \propto \cos \! \left( \varphi' \!\!\: - \varphi \right)$ as the probability of parallel motion is enhanced.
Thus, their product contains effectively positive \textit{polar alignment} terms proportional to~$\sin \! \left( \varphi' \!\!\: - \varphi \right)$, contributing to the first Fourier mode~$\int_{0}^{2\pi}d\varphi \sin \varphi \, \tilde{m}_2 \! \left( r,\alpha,\varphi \right)$ in the integral in Eq.~\eqref{eqn:orientation:tca}, even though polar alignment was not explicitly present  at the particle level.

The presence of correlations may renormalize the interaction parameters and even introduce new interaction terms at the field level.
Therefore, the coefficient~$\sigma_p$ can turn positive such that polar terms become relevant on the hydrodynamic level in a model with pure  nematic interactions.
These arguments crucially depend on the presence of self-propulsion which is the only term that breaks the global nematic symmetry.
Accordingly, polar order cannot emerge in the limit~$v_0 \rightarrow 0$.

\begin{figure}[t]
	\begin{center}
	\includegraphics[width=\columnwidth]{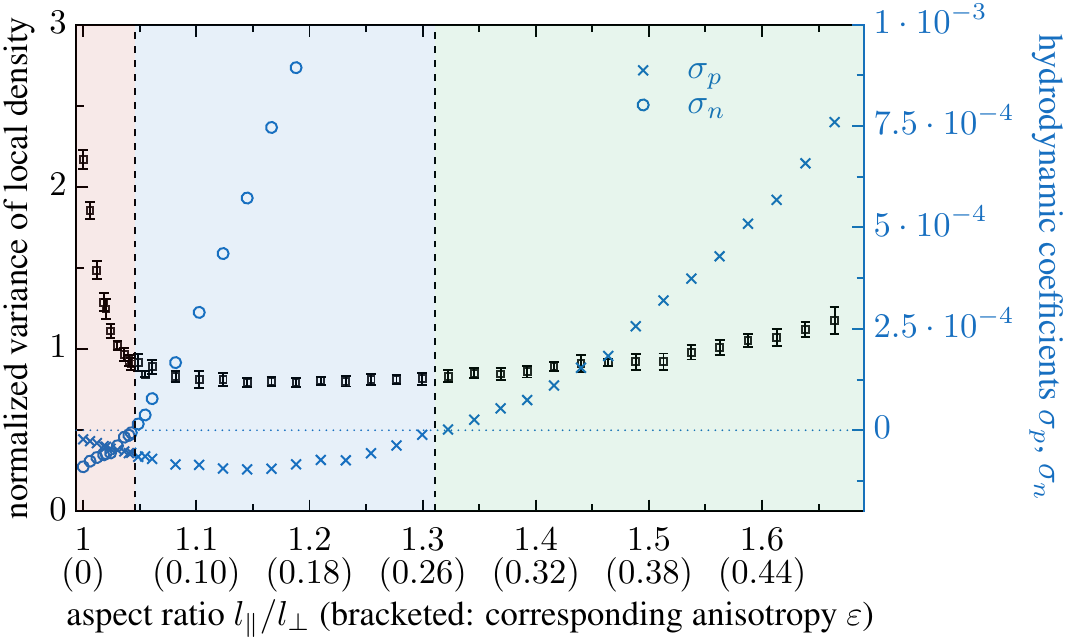}
	\end{center}
	\vspace{-0.6cm}
	\caption{From active phase separation  to the emergence  of different ordered regimes as a function of the anisotropy, for the parameters shown in Fig.~\ref{fig:mips_rods}. At first, MIPS breaks down as density fluctuations decrease, measured in terms of the variance of the coarse-grained density field, normalized with respect to the variance which is expected in a Poissonian point pattern as a proxy of a spatially homogeneous, disordered configuration. The first vertical line corresponds to the transition to local nematic order, signaled by~$\sigma_n=0$. The second vertical line indicates the onset of local polar order given by~$\sigma_p=0$ that leads to the formation of long-lived, large-scale polar clusters. The background color indicates the three identified phases/regimes; for typical snapshots, see Fig.~\ref{fig:mips_rods}. } 
	\label{fig:PhaseDiagram}
\end{figure}

We investigated numerically how the transport coefficients~$\sigma_p$ and $\sigma_n$ depend on the anisotropy of particles by measuring the pair distribution function~$g_2$ and evaluating the integrals in Eq.~\eqref{eqn:orientation:tc}. 
Figure~\ref{fig:PhaseDiagram} shows the relevant hydrodynamic coefficients together with the variance of the local density as a measure for the fluctuations of the particle density.
On the basis of this semi-analytical analysis, we can distinguish the following parameter ranges, which were introduced along with Fig.~\ref{fig:mips_rods} in \red{Section}~\ref{sec:pheno}.
Negative values of~$\sigma_p$ and~$\sigma_n$ together with a high level of density fluctuations reflects the properties of MIPS as observed for isotropic particles.
Increasing the anisotropy, density fluctuations decrease rapidly as MIPS aggregates break up, while~$\sigma_p$ and~$\sigma_n$ are negative~(local disorder).
Subsequently,~$\sigma_n$ turns positive signaling the emergence of local nematic order, followed by the emergence of local polar order when~$\sigma_p$ becomes  positive.

Thus, a theoretical description in terms of scalar quantities only, such as the particle density, is not applicable for  self-propelled rods once vectorial or tensorial order parameters grow at the local level:  the polar order parameter field~$\vec{p} \! \left( \vec{r},t \right)$ cannot be adiabatically eliminated if the hydrodynamic transport coefficient~$\sigma_p$ is positive.
In order to capture  anisotropic, nonequilibrium stresses on the coarse-grained level~--~arising due to the combined action of self-propulsion and anisotropic particle shape~--,~vectorial and tensorial order parameters have to be taken into account.
This underlines again the different nature between scalar and anisotropic active matter. 

\subsubsection{Discussion: symmetries of emergent ordered states}

%
%
In \red{Section}~\ref{sec:unfy_pat}, we reported the type of emergent patterns to be essentially dependent on the spatial dynamics, namely whether rods can possibly slip over each other or not.
To rationalize this observation, we consider the spatial dynamics~[Eq.~\eqref{eqn:spat_dyna}] comparing the order of magnitude of the active force propelling a particle forward with the characteristic magnitude of the repulsive force which is determined by the ratio of an energy scale~$\kappa$ and the typical size of a rod.
If the spatial dynamics is dominated by the active force, rods may easily slide over each other on a timescale which is faster than the timescale of the rotational dynamics, implying that the orientational dynamics is slow.
Therefore, one may simplify Eq.~\eqref{eqn:spat_dyna} to $\dot{\vec{r}}_k \simeq v_0 \vec{e} \! \left [ \varphi_k \right ]$.
In this limit, it turns out that the large-scale behaviour of spatially extended rods becomes similar to the dynamics of Vicsek-like rods with nematic alignment~\cite{ginelli_large_2010,shi_self_2018}.
Hence, the following phenomenology is expected:~for low rotational angular diffusion, a spatially homogeneous, nematic phase emerges at the mesoscale~(finite system size).
By increasing the noise, the level of nematic order decreases.
Close to the order-disorder transition, the system demixes into a high density region which is nematically ordered and a low-noise area where particles move in a disordered fashion.
This is the type of pattern shown on the right of Fig.~\ref{fig:NemaToPolarOrder}. 
We therefore conclude that the nematic alignment term in the torque [Eq. \eqref{eqn:bin_torques}] dominates the large-scale dynamics in the limit of high activity.
Accordingly, corresponding field theories for Vicsek-type self-propelled rods \cite{peshkov_nonlinear_2012,grossmann_mesoscale_2016} may account for the observed pattern formation phenomena, such as band formation and the emergence of nematic order.
Apparently, the fact that particles push each other is of minor importance in this parameter regime. 
In other words, positional correlations are less relevant if particles move fast as the system becomes well-mixed, i.e.~particle positions decorrelate quickly when rods can slip over each other.
%

%
By decreasing the self-propulsion force or, equivalently, increasing the repulsion strength, particles would get blocked upon encounter, positional correlations build up and the pair-correlation function becomes increasingly relevant such that mean-field arguments are not applicable. 
In this regime, patterns may emerge whose symmetry differs from the symmetries of the microscopic interaction.
It remains an open challenge for future work beyond the present study to derive the pair correlation function of anisotropic, self-propelled objects from first principles~--~analogues to corresponding theories for self-propelled discs~\cite{haertel_three_2018}~--~in order to explain the emergent bistability in parameter regimes where self-propulsion and repulsion are comparable which is most difficult to assess analytically as standard series expansions fail.

\section{Summary \& outlook}

We introduced a new approach to study active systems:~inspired by the quanto-mechanical wave-particle duality, active particles are represented by smooth fields and particle-particle interactions are derived from the minimization of an overlap energy.
Given the nature of the model both force and torque can be analytically obtained.
Importantly, this modeling technique allows studying the transition from self-propelled discs to active elongated, anisotropic, soft objects by performing continuous deformations on the shape as well as on the rigidity of the active particles.
Here, we analyzed analytically how motility-induced phase separated phases for isotropic, active particles become unstable for weakly anisotropic, self-propelled objects.
Specifically, we demonstrated that the combined action of anisotropic repulsion and self-propulsion leads to the emergence of effective torque terms which~--~above a critical particle aspect ratio~--~dissolve the polar boundary layer required to maintain motility-induced phase separated aggregates.
These findings provide an understanding of the role played by particle anisotropy regarding the robustness of the active phase separation described in terms of  scalar field theories for the particle density~(MIPS).

Furthermore, we used  the same theoretical framework to study the onset  of (orientationally) ordered phases.
We established that both particle aspect ratio and rigidity~--~in comparison with the self-propulsion~--~control the symmetry of the pair correlation function.
We demonstrated that the emerging order, either nematic or polar, is not dictated exclusively by the symmetry of the interacting potential but rather given by the resulting symmetry of both, interaction potential and the pair correlation function.
That is why polar and nematic structures can simultaneously coexist in a system of identical particles with a purely nematic interaction potential. 
This can be realized by fixing the rigidity and varying the particle aspect ratio, or fixing the aspect ratio and varying the rigidity.
At the level of the analytically derived coarse-grained order parameter equations, this would require showing  that it is possible to observe nematic order and vanishing polar order as well as non-vanishing polar and nematic order simultaneously.
In the latter scenario, the presence  of nonlinear cross-coupling terms in the evolution of  nematic and polar order parameters~\cite{peshkov_nonlinear_2012} implies that the magnitudes of these coefficients control whether nematic or polar orders prevail, or both coexist.
All these situations are possible for the studied system of anisotropic, self-propelled particles. 
In short, our analysis reveals that the symmetry of  macroscopic order is an emergent and dynamic property of active systems, as recently suggested from the analysis of a motility assay experiment~\cite{huber_emergence_2018}.

The developed framework allows studying motility-induced phase separation for isotropic, self-propelled particles, its breakup with particle anisotropy as well as the counterintuitive emergence of both, polar and nematic macroscopic order for the same type of active particles, thereby providing a unified picture of most relevant phenomena reported in active matter systems.  
We therefore expect our framework  to shed light on a large number of applications, including collective dynamics in collection of cells, the growth of bacterial colonies and self-organized patterns in systems of deformable, active filaments.  
Natural extensions of the developed approach range from the addition of hydrodynamic flows to the study of polydisperse systems, among many others.


\acknowledgments{
RG and FP  acknowledge support from the Agence Nationale de la Recherche via Grant No. ANR-15-CE30-0002-01. RG acknowledges funding from the People Programme (Marie Curie Actions) of the European Union's Seventh Framework Programme (FP7/2007-2013) under REA grant agreement n.~PCOFUND-GA-2013-609102, through the PRESTIGE programme coordinated by Campus France. ISA was supported by the 
NSF 446 PHY-1707900. }

\appendix

\section{Additional Langevin simulations}
\label{app:trans-pol-ord}

\paragraph*{Snapshots}

\begin{figure}[t]
 	\begin{center}
	\includegraphics[width=\columnwidth]{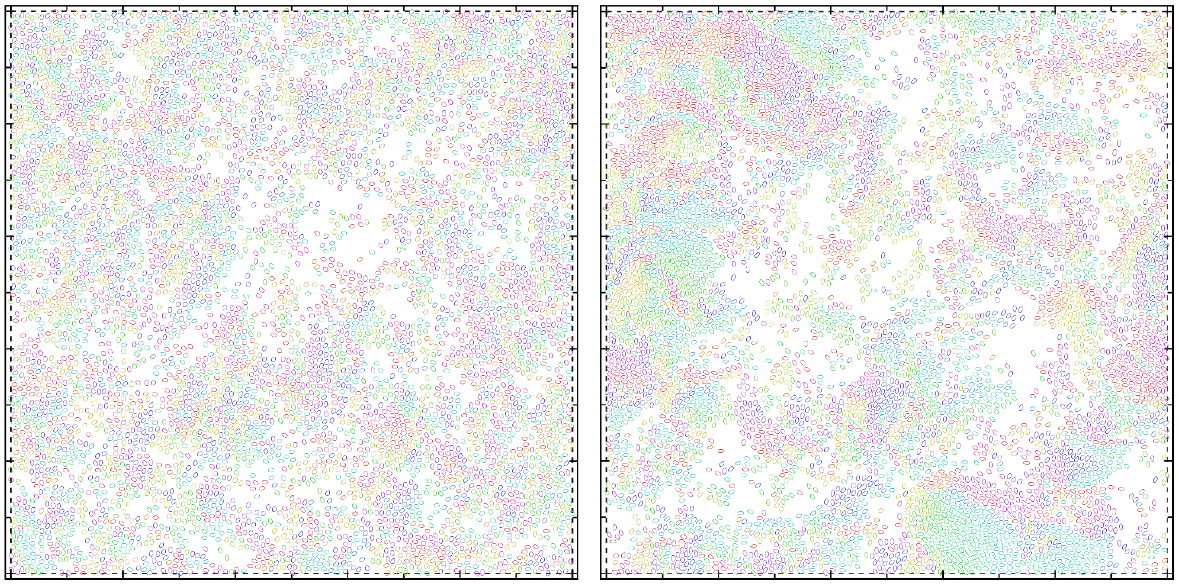}
 	\end{center}
	\vspace{-0.55cm}
	\caption{Growth of macroscopic polar domains for increasing anisotropies~$\varepsilon$, cf. to Fig.~\ref{fig:mips_rods}. The anisotropy is varied keeping the particle size $A \propto l_{\parallel} l_{\perp}$ fixed by setting the product $l_{\parallel} l_{\perp} = 1$. Left: $\varepsilon \approx 0.35 $, aspect ratio $l_{\parallel} / l_{\perp} = 1.44$; right: $\varepsilon \approx 0.48 $, aspect ratio $l_{\parallel} / l_{\perp} = 1.69$. Other parameters like in Fig.~\ref{fig:mips_rods}. \vspace*{0.6cm}}
	\label{fig:polartr}
\end{figure}

In addition to Fig.~\ref{fig:mips_rods}, the gradual emergence of local polar order for intermediate values of the anisotropy is illustrated in Fig.~\ref{fig:polartr}. In both cases, the transport coefficient~$\sigma_p$, which indicates the growth of local polar~[cf.~Eq.~\eqref{eqn:hydroP}] order, is positive.

\begin{figure}[t]
 	\begin{center}
 		\includegraphics[width=\columnwidth]{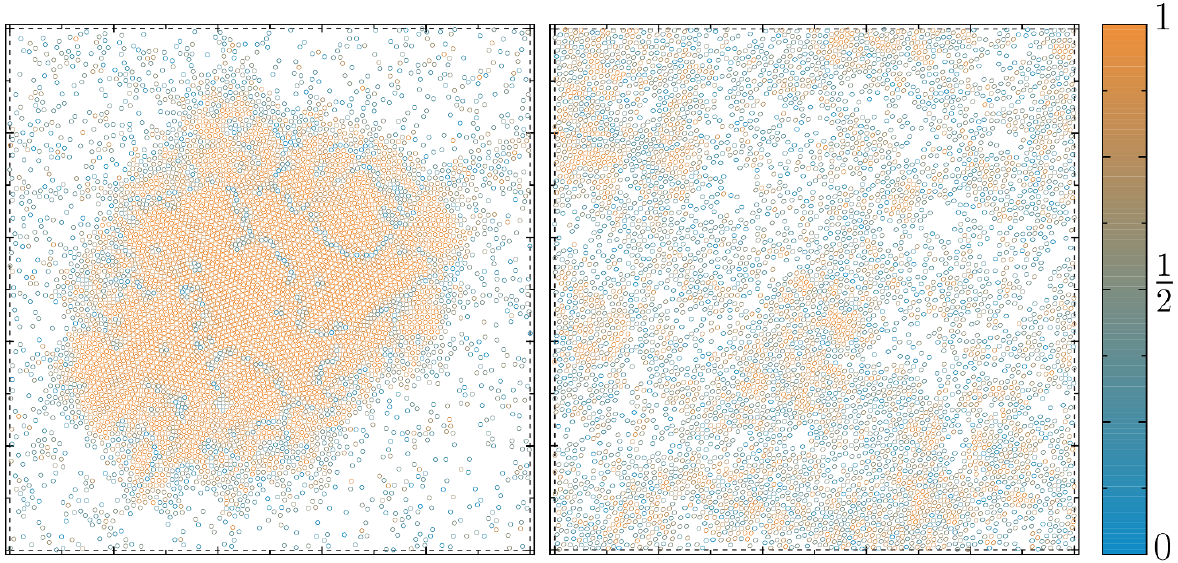}
 	\end{center}
	\vspace{-0.6cm}
	\caption{The same states like in the first two panels of Fig.~\ref{fig:mips_rods} are shown, however, the color coding corresponds to the absolute value of the local hexatic order parameter, defined via $\psi_6^{(k)} = \sum_{j \in \mathcal{V}_k} e^{6i \arg \left [ \vec{r}_j - \vec{r}_k \right ]} / \abs{\mathcal{V}_k}$, calculated for Voronoi neighbors $\mathcal{V}_k$~\cite{klein_algorithmisch_2005}. \textit{Left}: in the case of spheres ($\varepsilon = 0$), the system phase separates into a hexatic aggregate and a disordered gas; \textit{right}: for weak anisotropies of rods, the density fluctuations and the level of hexatic order are reduced due to anisotropic, active stresses. For parameters, see Figs.~\ref{fig:mips_rods}\red{(a,b)}. }
 	\label{fig:mips_rods:hex}
\end{figure}

The breakdown of a phase-separated state (MIPS) for anisotropic particles as shown in Fig.~\ref{fig:mips_rods} is accompanied with the breakdown of local hexatic order. Figure~\ref{fig:mips_rods:hex} shows the same states like Fig.~\ref{fig:mips_rods}, but the color coding corresponds to the local hexatic order parameter.

\paragraph*{Pair distribution functions}

\begin{figure}[b]
\begin{center}
\includegraphics[width=\columnwidth]{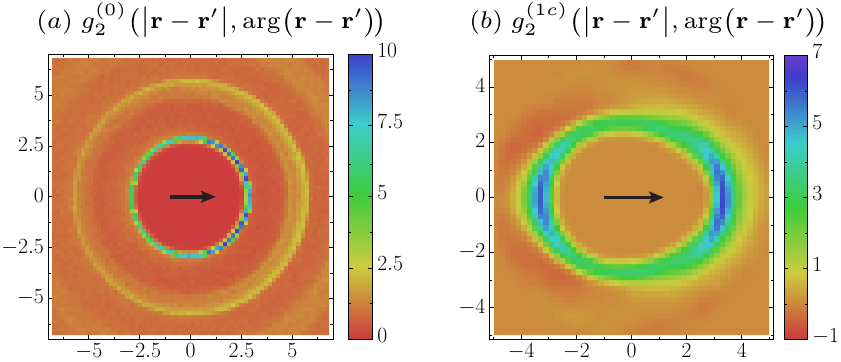}	
\vspace{-0.2cm}
\caption{(a) Pair correlation function~$g_2^{(0)}$ (color-coded) for spherical particles~($\varepsilon = 0$) in the disordered phase close to the MIPS transition. The correlation function is shown in the co-moving reference frame of a particle at the origin, moving towards the right as indicated by the black arrow. If the rotational diffusion  is low, the density distribution around a focal particle is asymmetric with respect to its direction of motion:~the probability to find a particle in front is significantly enhanced~\cite{bialke_microscopic_2013}. Parameters: energy functional $\mathcal{F} \! \left [ \xi \right ] = \xi^{\gamma}$ with energy scale $\kappa = 1$ and exponent $\gamma = 3$, speed $v_0 = 0.01$, translational and rotational mobilities $\mu_{\parallel} = \mu_{\perp} = 1$, $\mu_{\varphi} = 3/4$, translational  diffusion $D_{\parallel,\perp} = 0$, rotational diffusion  $D_{\varphi} = 7.5 \!\!\;\cdot \!\!\; 10^{-4}$,  systems size $L_{x,y} = 250$, particle number $N = 5968$. (b)
First Fourier component~$g_2^{(1c)}$ of the pair distribution function~$g_2$, indicating an enhanced probability of parallel motion of close-by rods due to occasional cluster formation.  The correlation functions further reveals an asymmetric excluded volume in contrast to self-propelled spheres, cf.~panel (a). Simulations were performed below the transition to local polar order~($\sigma_p < 0$). Parameters correspond to the simulation shown in Fig.~\ref{fig:mips_rods}\red{(c)}:~$\varepsilon \approx 0.19$ and $l_{\parallel} / l_{\perp} = 1.21$.}  
\label{fig:g2_spheres}
\end{center}
\end{figure}

A central theoretical argument is based on the enhancement of the probability to find a particle in front (with respect to the direction of self-propulsion) as compared to the probability to find it in its back.
This is reflected by the pair distribution function~$g_2$, specifically by the Fourier component
\begin{align}
	g_2^{(0)} \! & \left( \abs{\vec{r} - \vec{r}'} \! , \arg \! \left( \vec{r}' \! - \vec{r} \right) \right) \\
	& = \frac{1}{2\pi} \int_{-\pi}^{\pi} d\varphi \, g_2 \! \left( \abs{\vec{r} - \vec{r}'} \! , \arg \! \left( \vec{r}' \! - \vec{r} \right), \varphi \right) \nonumber
\end{align}
as shown in Fig.~\ref{fig:g2_spheres}~\red{(a)} for spherical particles, compare~\cite{bialke_microscopic_2013,haertel_three_2018}.

In \red{Section}~\ref{sec:em_ord}, we argued further that the probability of parallel motion is enhanced on average as a consequence of the anisotropic body shape~\cite{peruani_nonequilibrium_2006}, see also the schematic collisions in Fig.~\ref{fig:RodEx}.
This is also reflected by the pair correlation function.
In particular, a positive contribution to the Fourier component
\begin{align}
	g_2^{(1c)} \! & \left( \abs{\vec{r} - \vec{r}'} \! , \arg \! \left( \vec{r}' \! - \vec{r} \right) \right) \\
	& = \frac{1}{\pi} \int_{-\pi}^{\pi} d\varphi \, \cos \varphi \, g_2 \! \left( \abs{\vec{r} - \vec{r}'} \! , \arg \! \left( \vec{r}' \! - \vec{r} \right), \varphi \right) \nonumber
\end{align}
reflects that the probability of moving together in groups ($\varphi \approx \varphi'$) is larger than swimming in an anti-parallel fashion~($\varphi \approx \varphi + \pi$).
This argument has been verified by numerical measurements of the respective part of the pair distribution function, cf.~Fig.~\ref{fig:g2_spheres}\red{(b)}.

\paragraph*{Movies of Langevin simulations}

We provide three movies~\cite{SI} for the nontrivial states shown in Fig.~\ref{fig:mips_rods}: mov1.mp4 (\ref{fig:mips_rods}\red{a}), mov2.mp4 (\ref{fig:mips_rods}\red{d}) and mov3.mp4 (\ref{fig:mips_rods}\red{e}). 
Furthermore, the animation mov4.mp4 corresponds to the simulation which is depicted in Fig.~\ref{fig:NemaToPolarOrder}\red{(b)}.

\newpage
\clearpage


%

\end{document}